\documentclass[a4paper,11pt]{article}
\usepackage{float}
\usepackage{jheppub}
\usepackage{epsfig}
\usepackage[T1]{fontenc}
\def \be {\begin{equation}}
\def \ee {\end{equation}}
\def \bea {\begin{eqnarray}}
\def \eea {\end{eqnarray}}

\title{\boldmath The trace anomaly for a chiral fermion}

\author{Chang-Yong Liu}

\affiliation{College of Science, Northwest A\&F University, Yangling, Shaanxi 712100, China}

\emailAdd{liuchangyong@nwsuaf.edu.cn}

\abstract{In this paper, we study the problem of trace anomaly for a chiral fermion. To find whether there exists a parity-odd term (Pontryagin term), we use a modified Breitenlohner-Maison-'t Hooft-Veltman regularization and Fujikawa's method by a new Dirac mass term to calculate the trace anomaly for a Weyl fermion coupled to an abelian gauge field and a gravity separately. We show once again that the trace anomaly of energy-momentum tensor for a chiral fermion has Pontryagin term.}


\arxivnumber{}

\begin{document}
\maketitle
\section{Introduction}

In particle physics, symmetry are some of the most important concepts for the theory to be renormalizable and unitary. If the classical symmetry of the Lagrangian cannot be maintained in the process of quantization, the theory is said to have an anomaly. Anomaly originated from the research on the decay of the neutral pion which can be well understood by chiral anomaly (Adler-Bell-Jackiw anomaly \cite{Adler:1969gk,Bell:1969ts}). The chiral anomalies have various applications in particle physics, e.g. \cite{Gross:1972pv,Liu:2021pmt,Liu:2022zdo}. There are many other types of anomaly: gravitational anomaly \cite{Alvarez-Gaume:1983ihn}, trace anomaly \cite{Capper:1974ic,Capper:1975ig,Deser:1976yx,Christensen:1976vb,Duff:1977ay,Dowker:1976zf,Wald:1978pj,Bonora:1985cq,Bonora:1984ic,Duff:1993wm}. Among these anomalies, the most
mysterious one is the trace anomaly for a Weyl fermion. The different regularizing schemes have led to some controversial results whose difference focus on the the parity-odd term.  Suggested by the work \cite{Nakayama:2012gu}, the group of Bonora et al. has claimed that there exists a parity-odd term in the trace anomaly of Weyl fermions \cite{Bonora:2014qla,Bonora:2015nqa,Bonora:2017gzz,Bonora:2018obr,Bonora:2020upc}. The result is confirmed by the Pauli-Villars regularization in \cite{Liu:2022jxz}. There is other opposite conclusion that the parity-odd term do not arise at all by the group of Bastianelli et al. \cite{Bastianelli:2016nuf,Bastianelli:2018osv,Bastianelli:2019fot,Bastianelli:2019zrq,Bastianelli:2022hmu},  Fr\"{o}b et al. \cite{Frob:2019dgf} and Abdallah et al. \cite{Abdallah:2021eii}. The ambiguity arises from that the trace anomaly is unrelated to diffeomorphisms and other symmetries of the theory. The accidental vanishing of the odd three-point function of the energy-momentum tensor \cite{Bonora:2015nqa} leads to the different results \cite{Bonora:2022izj}.

In this letter, we continue to study the trace anomaly by dimensional regularization and Fujikawa's method separately. Though the Breitenlohner-Maison-'t Hooft-Veltman (BMHV) regularization \cite{tHooft:1972tcz,Breitenlohner:1977hr,Breitenlohner:1975hg,Breitenlohner:1976te} give mathematically consistent results to arbitrary loop orders for non-chiral theories \cite{Breitenlohner:1977hr,Rufa:1990hg}, it can not apply to chiral theories directly. We will use a modified BMHV regularization to calculate the trace anomaly for a Weyl fermion coupled to an abelian gauge field. To confirm these results, we also use the Fujikawa's method \cite{Fujikawa:1979ay,Fujikawa:1980vr} with a new Dirac mass term for Pauli-Villars field \cite{Pauli:1949zm} to calculate it. The paper is organized as follows. In the next section, we use a modified BMHV regularization to calculate the trace anomaly for a Weyl fermion coupled to an abelian gauge field. In Section 3, we apply the Fujikawa's method to evaluate the trace anomaly by Pauli-Villars regularization. We end with the conclusions. Some definitions and useful formulae are put in appendix.\\

\section{Dimensional regularization}
In this section, we use a modified BMHV regularization to calculate the trace anomaly for a Weyl fermion. We consider the Lagrangian for a left-handed Weyl fermion coupled to an abelian gauge field $A_{\mu}$ to be
\bea
\label{Lagrangian}
\mathcal{L}=-\overline{\psi_L}D\!\!\!\!/\psi_L=-\overline{\psi_L}D\!\!\!\!/(A)\psi_L
\eea
with $\psi_L=P_L\psi$ and $D\!\!\!\!/=\gamma^{\mu}D_{\mu}$, where $\psi$ is Dirac field and $D_{\mu}=\partial_{\mu}-iA_{\mu}$ is the gauge covariant derivative.
For later use, we couple the theory to gravity by introducing the vierbein ${e_{\mu}}^a$ ($\mu,\nu,\cdots$ are world indices, $a,b,\cdots$ are flat indices). The full coupling to gravity Lagrangian $\mathcal{L}^{(g)}$ is
\bea
\label{Lagrangiangravity}
\mathcal{L}^{(g)}=-e\overline{\psi_L}\gamma^{\mu}\nabla_{\mu}\psi_L,
\eea
where $\gamma^{\mu}={e^{\mu}}_a\gamma^a$ are the gamma matrices with curved indices and $e$ is the determinant of the vierbein. The $\nabla_{\mu}$ is the covariant derivative containing the spin connection $\omega_{\mu a b}$
\bea
\label{covariant}
\nabla_{\mu}=\partial_{\mu}-iA_{\mu}+\frac{1}{4}\omega_{\mu a b}\gamma^a \gamma^b.
\eea

The action $S=\int d^4x \mathcal{L}^{(g)}$
is gauge invariant which can be written as
\bea
\label{u1}
\begin{cases}
   \psi_L(x) \quad \rightarrow \quad \psi_L'(x)={\rm{e}}^{i \alpha(x)}\psi_L(x)\\
    \overline{\psi_L}(x)\quad \rightarrow \quad \overline{\psi_L}'(x)={\rm{e}}^{-i \alpha(x)}\overline{\psi_L}(x)\\
    A_{\mu}(x) \quad \rightarrow \quad A_{\mu}'(x)=A_{\mu}(x)+\partial_{\mu} \alpha(x)\\
    e^{\ a}_{\mu}(x)\quad \rightarrow \quad e'^{\ a}_{\mu}(x)=e^{\ a}_{\mu}(x).
 \end{cases}
\eea
The corresponding gauge current in the flat space limit is given by
\bea
j^{\mu}=i\overline{\psi_L}\gamma^{\mu}\psi_L.
\eea
The local Weyl symmetry for the action $S$ is given by
\bea
\label{weyl}
\begin{cases}
   \psi_L(x) \quad \rightarrow \quad \psi_L'(x)={\rm{e}}^{-\frac{3}{2} \sigma(x)}\psi_L(x)\\
    \overline{\psi_L}(x)\quad \rightarrow \quad \overline{\psi_L}'(x)={\rm{e}}^{-\frac{3}{2} \sigma(x)}\overline{\psi_L}(x)\\
    A_{\mu}(x) \quad \rightarrow \quad A_{\mu}'(x)=A_{\mu}(x)\\
    e^{\ a}_{\mu}(x)\quad \rightarrow \quad e'^{\ a}_{\mu}(x)={\rm{e}}^{\sigma(x)}e^{\ a}_{\mu}(x).
 \end{cases}
\eea
The stress tensor is defined by
\bea
T^{\mu a}(x)=\frac{1}{e}\frac{\delta S}{\delta e_{\mu a}(x)}.
\eea
In the flat space limit, the stress tensor has the form
\bea
\label{stresstensor}
T^{\mu\nu}=\frac{1}{4}\overline{\psi_L}(\gamma^{\mu}\overleftrightarrow{D}^{\nu}+\gamma^{\nu}\overleftrightarrow{D}^{\mu})\psi_L.
\eea
Where $\overleftrightarrow{D}^{\mu}=D^{\mu}-\overleftarrow{D}^{\mu}$ and $\overleftarrow{D}^{\mu}=\overleftarrow{\partial}^{\mu}+iA^{\mu}$.
The stress tensor (\ref{stresstensor}) is traceless on-shell
\bea
T^{\mu}_{\  \ \mu}=0
\eea
and satisfies the equation
\bea
\label{classicalstresstensor}
\partial^{\mu}T_{\mu\nu}=-j^{\mu}F_{\mu\nu}.
\eea
The global chiral rotation
\bea
\label{globalchiral}
\begin{cases}
   \psi_L(x) \quad \rightarrow \quad \psi_L'(x)={\rm{e}}^{i \beta\gamma^5}\psi_L(x)\\
    \overline{\psi_L}(x)\quad \rightarrow \quad \overline{\psi_L}'(x)=\overline{\psi_L}(x){\rm{e}}^{i \beta\gamma^5}\\
    A_{\mu}(x) \quad \rightarrow \quad A_{\mu}'(x)=A_{\mu}(x)\\
    e^{\ a}_{\mu}(x)\quad \rightarrow \quad e'^{\ a}_{\mu}(x)=e^{\ a}_{\mu}(x),
 \end{cases}
\eea
with constant $\beta$, is a symmetry of the Lagrangian (\ref{Lagrangiangravity}). The axial vector current in the flat space limit is
\bea
j^{\mu 5}=i\overline{\psi_L}\gamma^{\mu}\gamma^5\psi_L=i\overline{\psi_L}\gamma^{\mu}\psi_L=j^{\mu}.
\eea
Then classically the axial vector current $j^{\mu 5}$ has the same expression as the gauge current $j^{\mu}$ in chiral theory.

The trace anomaly for a left-handed Weyl fermion has been studied by the BMHV prescription in works \cite{Bonora:2020upc,Bastianelli:2022hmu}. There are some problems in BMHV prescription to deal with the chiral fields \cite{Jegerlehner:2000dz}. It is difficult to regularize fermion loop integral by continuation in $n$ dimensions. To see this, we
continue the Lagrangian (\ref{Lagrangian}) from the $4$ to $n$ dimensions, that is
\bea
\mathcal{L}^{(n)}=-\overline{\psi}\dot{P}_RD\!\!\!\!/\dot{P}_L\psi=-\overline{\psi}\dot{P}_R\dot{\gamma} D_{\mu}\dot{P}_L\psi=-\overline{\psi}\dot{P}_R\overline{\gamma} \overline{D}_{\mu}\dot{P}_L\psi=\mathcal{L}^{(4)}.
\eea
The $\mathcal{L}^{(n)}$ is the $n$ dimensional Lagrangian and $\mathcal{L}^{(4)}$ is the $4$ dimensional one. So the Lagrangian in $n$ dimensions is the same as the one in $4$ dimensions. To overcome this problem, it is useful to rewrite the Lagrangian (\ref{Lagrangian}) in terms of a non-chiral fermion field $\psi$ by inserting one chirality projector:
\bea
\label{Lagrangianooo}
\mathcal{\dot{L}}^{(n)}=-\bar{\psi} D\!\!\!\!/ \dot{P}_L \psi.
\eea
The Lagrangian is useful in studying the anomalies \cite{Weinberg:1996kr,Bilal:2008qx}. In \cite{Bonora:2020upc}, the authors find that the trace anomaly has parity-odd term with this Lagrangian (\ref{Lagrangianooo}). There is another problem in ordinary BMHV prescription that the Ward identity associated with the conservation of the stress tensor (\ref{classicalstresstensor}) is not satisfied. We will come to this point in the end of this section. Thus we adopt a modified BMHV scheme.

\begin{figure}[H]
\center
\includegraphics[width=0.65\textwidth]{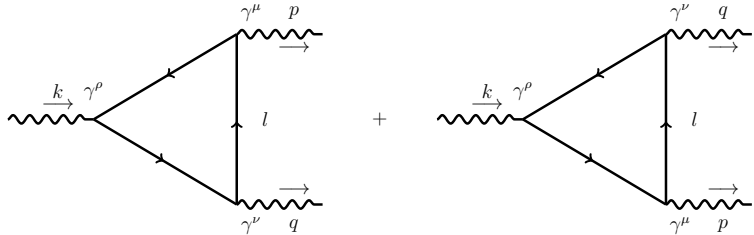}
\caption{\label{triangle} Triangle diagrams contributing to chiral anomaly.  }
\end{figure}

\begin{figure}[H]
\center
\includegraphics[width=0.65\textwidth]{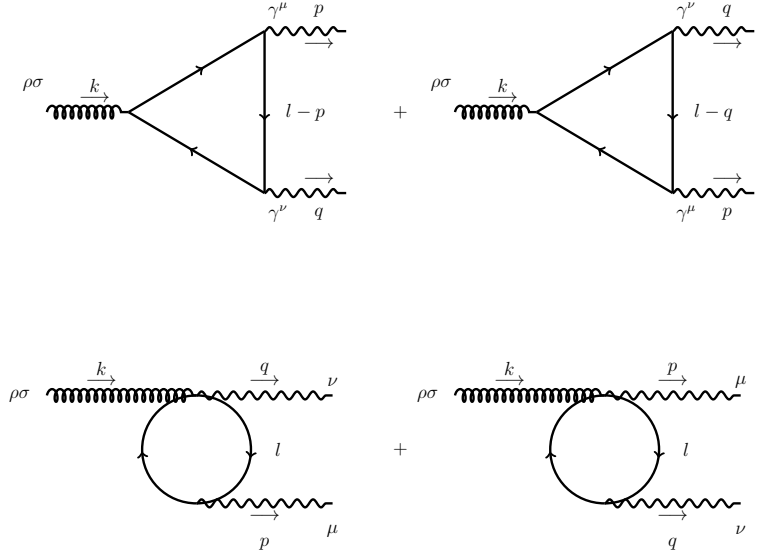}
\caption{\label{tracefigure} The amplitudes contributing to the stress tensor amplitude (\ref{expansion23}).  }
\end{figure}

A modified BMHV scheme is described as follow: we first use the naive dimensional regularization (see appendix A) to express the amplitudes with only one projectors $P_L$ or $P_R$ left and then apply the ordinary BMHV scheme to obtain the final result. In this scheme, the Lagrangian $\mathcal{\ddot{L}}^{(n)}$ have the equivalent forms
\bea
\mathcal{\ddot{L}}^{(n)}=-\overline{\psi_L}D\!\!\!\!/\psi_L=-\overline{\psi_L}D\!\!\!\!/(A)\psi_L=-\bar{\psi}P_R D\!\!\!\!/ P_L \psi=-\bar{\psi} D\!\!\!\!/ P_L \psi=-\bar{\psi} P_R D\!\!\!\!/ \psi.
\eea
Then the fermion propagator and vertex are
\bea
P \quad &:& \quad P_L\frac{-p\!\!\!/}{p^2-i\epsilon}=\frac{-p\!\!\!/}{p^2-i\epsilon}P_R=p_L\frac{-p\!\!\!/}{p^2-i\epsilon}P_R\nonumber\\
V \quad &:& \quad -\gamma^{\mu} P_L=-P_R\gamma^{\mu}=-P_R\gamma^{\mu}P_L.
\eea
We now study the chiral and trace anomaly by this modified BMHV scheme. In the formalism of the background fields, the chiral anomaly can be obtained from the functional integral
\bea
\label{expansion}
\left\langle j^{\rho}(x)\right\rangle_A&=&\int \mathcal{D}\psi\mathcal{D}\overline{\psi}j^{\rho}(x) {\rm{e}}^{i \int d^4y\mathcal{L}+i \int d^4y j^{\lambda}(y)A_{\lambda}(y)}\nonumber\\
&=&\left\langle j^{\rho}(x){\rm{e}}^{i \int d^4y j^{\lambda}(y)A_{\lambda}(y)}\right\rangle \nonumber\\
&=&-\frac{1}{2}\int d^4 y \int d^4 z A_{\mu}(y)A_{\nu}(z)\left\langle j^{\rho}(x)j^{\mu}(y)j^{\nu}(z)\right\rangle+\ldots \nonumber\\
&=&-\frac{1}{2}\int d^4 y \int d^4 z A_{\mu}(y)A_{\nu}(z)\int \frac{d^4 k}{(2\pi)^4}\int \frac{d^4 p}{(2\pi)^4}\int \frac{d^4 q}{(2\pi)^4}\nonumber\\
&&\times {\rm{e}}^{-ikx}{\rm{e}}^{ipy}{\rm{e}}^{iqz}(2\pi)^4 \delta^{(4)}(k-p-q)\mathcal{M}^{\rho \mu \nu}(p,q)+\ldots.
\eea
The expression of the triangle diagrams is (Figure \ref{triangle})
\bea
\label{trianglediagrams}
\mathcal{M}^{\rho \mu \nu}(p,q)=i\int \frac{d^n l}{(2 \pi)^n}\frac{{\rm{tr}}[\gamma^{\rho} P_L (l\!\!\!/-p\!\!\!/)\gamma^{\mu}P_L l\!\!\!/ \gamma^{\nu}P_L (l\!\!\!/+q\!\!\!/)]}{l^2(l-p)^2(l+q)^2}+\bigg((p,\mu) \leftrightarrow (q,\nu)\bigg).
\eea
The chiral anomaly can be obtained through second order in the external field $A$. The divergence in momentum space is equivalent to
\bea
-ik_{\rho}\mathcal{M}^{\rho \mu \nu}&=&\int \frac{d^n l}{(2 \pi)^n}\frac{{\rm{tr}}[k\!\!\!/ P_L (l\!\!\!/-p\!\!\!/)\gamma^{\mu}P_L l\!\!\!/ \gamma^{\nu}P_L (l\!\!\!/+q\!\!\!/)]}{l^2(l-p)^2(l+q)^2}+\bigg((p,\mu) \leftrightarrow (q,\nu)\bigg)\nonumber\\
&=& \int \frac{d^n l}{(2 \pi)^n}\frac{{\rm{tr}}[k\!\!\!/  (l\!\!\!/-p\!\!\!/)\gamma^{\mu} l\!\!\!/ \gamma^{\nu} (l\!\!\!/+q\!\!\!/)\boldsymbol{P_R}]}{l^2(l-p)^2(l+q)^2}+\bigg((p,\mu) \leftrightarrow (q,\nu)\bigg)
\eea
We set the external momentums to be $p^2=q^2=0$ and apply the BMHV scheme to get
\bea
\label{case1}
-ik_{\rho}\mathcal{M}^{\rho \mu \nu}&=&\int \frac{d^n l}{(2 \pi)^n}\frac{{\rm{tr}}[k\!\!\!/  (l\!\!\!/-p\!\!\!/)\gamma^{\mu} l\!\!\!/ \gamma^{\nu} (l\!\!\!/+q\!\!\!/)\boldsymbol{\dot{P}_R}]}{l^2(l-p)^2(l+q)^2}+\bigg((p,\mu) \leftrightarrow (q,\nu)\bigg) \nonumber\\
&=&-\frac{1}{4\pi^2}\epsilon^{\mu\nu\rho\sigma}p_{\rho}q_{\sigma},
\eea
From the expansion (\ref{expansion}), we have
\bea
\left\langle \partial_{\rho}j^{\rho}(x)\right\rangle&=&\frac{i}{2}\int d^4 y \int d^4 z A_{\mu}(y)A_{\nu}(z)\int \frac{d^4 k}{(2\pi)^4}\int \frac{d^4 p}{(2\pi)^4}\int \frac{d^4 q}{(2\pi)^4}\nonumber\\
&&\times {\rm{e}}^{-ikx}{\rm{e}}^{ipy}{\rm{e}}^{iqz}(2\pi)^4 \delta^{(4)}(k-p-q)k_{\rho}\mathcal{M}^{\rho \mu \nu}(p,q)\nonumber\\
&=&\frac{1}{32\pi^2}\epsilon^{\mu\nu\rho\sigma}F_{\mu\nu}(x)F_{\rho\sigma}(x).
\eea
There are other places that the $P_L$ or $P_R$ can be located at. For example, the triangle diagrams (\ref{trianglediagrams}) has the equivalent form
\bea
\mathcal{M}^{\rho \mu \nu}(p,q)=i\int \frac{d^n l}{(2 \pi)^n}\frac{{\rm{tr}}[\gamma^{\rho}  (l\!\!\!/-p\!\!\!/)\gamma^{\mu} l\!\!\!/ \gamma^{\nu}\boldsymbol{P_L} (l\!\!\!/+q\!\!\!/)]}{l^2(l-p)^2(l+q)^2}+\bigg((p,\mu) \leftrightarrow (q,\nu)\bigg).
\eea
After we apply the BMHV scheme
\bea
\hat{\mathcal{M}}^{\rho \mu \nu}(p,q)=i\int \frac{d^n l}{(2 \pi)^n}\frac{{\rm{tr}}[\gamma^{\rho}  (l\!\!\!/-p\!\!\!/)\gamma^{\mu} l\!\!\!/ \gamma^{\nu}\boldsymbol{\dot{P}_L} (l\!\!\!/+q\!\!\!/)]}{l^2(l-p)^2(l+q)^2}+\bigg((p,\mu) \leftrightarrow (q,\nu)\bigg),
\eea
we obtain
\bea
\label{case2}
\begin{cases}
  -ik_{\rho}\hat{\mathcal{M}}^{\rho \mu \nu}=0;\\
    -ip_{\mu}\hat{\mathcal{M}}^{\rho \mu \nu}=-\frac{2i\pi^2(n-2)}{n-1}B_0(k^2,0,0)[k^{\rho}k^{\nu}-g^{\rho\nu}k^2];\\
    -iq_{\nu}\hat{\mathcal{M}}^{\rho \mu \nu}=-\frac{1}{4\pi^2}\epsilon^{\mu\rho\nu\sigma}p_{\nu}q_{\sigma}+\frac{2i\pi^2(n-2)}{n-1}B_0(k^2,0,0)[k^{\rho}k^{\mu}-g^{\mu\rho}k^2].
  \end{cases}
\eea
Where the function $B_0(x,0,0)$ is
\bea
\label{B000}
B_0(x,0,0)=\frac{1}{8\pi^4(4-n)}+\frac{\log(-\frac{4\pi \mu^2}{x})}{16\pi^4}-\frac{\gamma-2}{16\pi^4}+O(4-n)
\eea
and the $\gamma$ is the Euler-Mascheroni constant. The result (\ref{case2}) is different with the one (\ref{case1}), so we have some freedom in deciding which current exhibits the anomaly. The evaluation of the anomalies depends on the choice we make of the place $P_L$ or $P_R$. On the other hand, the form of the chiral anomaly which satisfies the Wess-Zumino consistency condition is independent of the details of the theory. The chiral anomaly are given by non-trivial BRST cohomology classes at ghost number one on the space of local functionals \cite{Weinberg:1996kr,Bilal:2008qx}. The anomaly can move from one current to others by the addition of a local counterterm to the classical action. The place $P_L$ or $P_R$ need to be in accordance with the special features of the problem at hand, so we require that the gauge current $j^{\mu}$ which couples to external gauge field $A_{\mu}$ is free of anomaly in expansion (\ref{expansion}). To distinguish the gauge current which free of anomaly, the axial vector current $j^{\rho5}(x)$ need to have chiral anomaly. This criterion is the same as the calculation of the expectation value of stress tensor (\ref{stresstensor}) which can be expanded in the form
\bea
\label{expansion2}
\left\langle T^{\rho\sigma}(x)\right\rangle_A&=&\int \mathcal{D}\psi\mathcal{D}\overline{\psi}T^{\rho\sigma}(x) {\rm{e}}^{i \int d^4y\mathcal{L}+i \int d^4y j^{\lambda}(y)A_{\lambda}(y)}\nonumber\\
&=&\left\langle T^{\rho\sigma}(x){\rm{e}}^{i \int d^4y j^{\lambda}(y)A_{\lambda}(y)}\right\rangle \nonumber\\
&=&-\frac{1}{2}\int d^4 y \int d^4 z A_{\mu}(y)A_{\nu}(z)\left\langle T^{\rho\sigma}(x)j^{\mu}(y)j^{\nu}(z)\right\rangle+i\int d^4 y  A_{\mu}(y)\left\langle T^{\rho\sigma}(x)j^{\mu}(y)\right\rangle+\ldots \nonumber\\
&=&-\frac{1}{2}\int d^4 y \int d^4 z A_{\mu}(y)A_{\nu}(z)\int \frac{d^4 k}{(2\pi)^4}\int \frac{d^4 p}{(2\pi)^4}\int \frac{d^4 q}{(2\pi)^4}\nonumber\\
&&\times {\rm{e}}^{-ikx}{\rm{e}}^{ipy}{\rm{e}}^{iqz}(2\pi)^4 \delta^{(4)}(k-p-q)T^{\rho \sigma \mu \nu}(p,q)+\ldots.
\eea
Where the $T^{\rho\sigma\mu\nu}(p,q)$ is (Figure \ref{tracefigure})
\bea
\label{expansion23}
T^{\rho\sigma\mu\nu}(p,q)&=&-\frac{i}{4}\int \frac{d^n l}{(2 \pi)^n}\frac{N^{\rho\sigma\mu\nu}}{l^2(l-p)^2(l-p-q)^2}\nonumber\\
&+&\frac{i}{2}\int \frac{d^n l}{(2 \pi)^n}\frac{M^{\rho\sigma\mu\nu}}{l^2(l-p)^2}
+\bigg((p,\mu) \leftrightarrow (q,\nu)\bigg).
\eea
The numerators $N^{\rho\sigma\mu\nu}$ and $M^{\rho\sigma\mu\nu}$ are
\bea
&&N^{\rho\sigma\mu\nu}={\rm{tr}}\big[\big((2l-p-q)^{\rho}\gamma^{\sigma}+(2l-p-q)^{\sigma}\gamma^{\rho} \big)P_L l\!\!\!/\gamma^{\mu}P_L (l\!\!\!/-p\!\!\!/) \gamma^{\nu}P_L (l\!\!\!/-p\!\!\!/-q\!\!\!/) \big],\nonumber\\
&&M^{\rho\sigma\mu\nu}={\rm{tr}}\big[\big(\gamma^{\rho}g^{\nu\sigma}+\gamma^{\sigma}g^{\nu\rho} \big)P_L l\!\!\!/\gamma^{\mu}P_L (l\!\!\!/-p\!\!\!/) \big].
\eea
This leads to
\bea
T^{\mu\nu}(p,q)=g_{\rho\sigma}T^{\rho\sigma\mu\nu}(p,q)&=&-\frac{i}{2}\int \frac{d^n l}{(2 \pi)^n}\frac{{\rm{tr}}\big[(2l\!\!\!/-p\!\!\!/-q\!\!\!/)P_L l\!\!\!/\gamma^{\mu}P_L (l\!\!\!/-p\!\!\!/) \gamma^{\nu}P_L (l\!\!\!/-p\!\!\!/-q\!\!\!/)\big]}{l^2(l-p)^2(l-p-q)^2}\nonumber\\
&+&i\int \frac{d^n l}{(2 \pi)^n}\frac{{\rm{tr}}\big[\gamma^{\nu}P_L l\!\!\!/\gamma^{\mu}P_L (l\!\!\!/-p\!\!\!/) \big]}{l^2(l-p)^2}+\bigg((p,\mu) \leftrightarrow (q,\nu)\bigg).
\eea
The gauge current $j^{\mu}$ which couples to external gauge field in the expansion (\ref{expansion2}) is free of anomaly, then the $T^{\mu\nu}(p,q)$ has the form
\bea
T^{\mu\nu}(p,q)&=&-\frac{i}{2}\int \frac{d^n l}{(2 \pi)^n}\frac{{\rm{tr}}\big[(2l\!\!\!/-p\!\!\!/-q\!\!\!/) l\!\!\!/\gamma^{\mu} (l\!\!\!/-p\!\!\!/) \gamma^{\nu} (l\!\!\!/-p\!\!\!/-q\!\!\!/) \dot{P}_R \big]}{l^2(l-p)^2(l-p-q)^2}\nonumber\\
&+&i\int \frac{d^n l}{(2 \pi)^n}\frac{{\rm{tr}}\big[\gamma^{\nu}l\!\!\!/\gamma^{\mu}\dot{P}_L (l\!\!\!/-p\!\!\!/) \big]}{l^2(l-p)^2}+\bigg((p,\mu) \leftrightarrow (q,\nu)\bigg)\nonumber\\
&=&\frac{i}{8\pi^2}\epsilon^{\mu\nu\rho\sigma}p_{\rho}q_{\sigma}.
\eea
Then the parity-odd term of trace anomaly $\left\langle T^{\rho}_{ \  \rho}(x)\right\rangle^{odd}$ is
\bea
\left\langle T^{\rho}_{ \  \rho}(x)\right\rangle^{odd}&=&-\frac{1}{2}\int d^4 y \int d^4 z A_{\mu}(y)A_{\nu}(z)\int \frac{d^4 k}{(2\pi)^4}\int \frac{d^4 p}{(2\pi)^4}\int \frac{d^4 q}{(2\pi)^4}\nonumber\\
&&\times {\rm{e}}^{-ikx}{\rm{e}}^{ipy}{\rm{e}}^{iqz}(2\pi)^4 \delta^{(4)}(k-p-q)T^{\mu\nu}(p,q)\nonumber\\
&=&-\frac{i}{64\pi^2}\epsilon^{\mu\nu\rho\sigma}F_{\mu\nu}(x)F_{\rho\sigma}(x).
\eea

The trace anomaly also has the parity-even part $\left\langle T^{\rho}_{ \  \rho}(x)\right\rangle^{even}$ which has the form \cite{Peskin:1995ev}
\bea
\label{parityeven}
\left\langle T^{\rho}_{ \  \rho}(x)\right\rangle^{even}=\frac{\beta(e)}{2e^3}F_{\mu\nu}(x)F^{\mu\nu}(x).
\eea
The $\beta$ function can be obtained by standard method. As the vacuum polarization of a Weyl fermion in modified BMHV scheme is half the one of a Dirac fermion, i.e.
\bea
\label{Weydirac}
\Pi^{\mu\nu}(p)=\int \frac{d^n l}{(2 \pi)^n}\frac{{\rm{tr}}\big[\gamma^{\mu}P_L l\!\!\!/\gamma^{\nu}P_L (l\!\!\!/-p\!\!\!/) \big]}{l^2(l-p)^2}&=&\int \frac{d^n l}{(2 \pi)^n}\frac{{\rm{tr}}\big[\gamma^{\mu}l\!\!\!/\gamma^{\nu} \dot{P}_L(l\!\!\!/-p\!\!\!/) \big]}{l^2(l-p)^2} \nonumber\\
&=&\frac{1}{2}\int \frac{d^n l}{(2 \pi)^n}\frac{{\rm{tr}}\big[\gamma^{\mu}l\!\!\!/\gamma^{\nu} (l\!\!\!/-p\!\!\!/) \big]}{l^2(l-p)^2},
\eea
the $\beta(e)$ function is
\bea
\beta(e)=\frac{e^3}{24\pi^2}.
\eea
Substituting this into (\ref{parityeven}), we have
\bea
\left\langle T^{\rho}_{ \  \rho}(x)\right\rangle^{even}=\frac{1}{48\pi^2}F_{\mu\nu}(x)F^{\mu\nu}(x).
\eea
Then the trace anomaly is
\bea
\label{traceanomaly69}
\left\langle T^{\rho}_{ \  \rho}(x)\right\rangle&=&\left\langle T^{\rho}_{ \  \rho}(x)\right\rangle^{odd}+\left\langle T^{\rho}_{ \  \rho}(x)\right\rangle^{even}\nonumber \\
&=&-\frac{i}{64\pi^2}\epsilon^{\mu\nu\rho\sigma}F_{\mu\nu}(x)F_{\rho\sigma}(x)-\frac{1}{48\pi^2}F_{\mu\nu}(x)F^{\mu\nu}(x).
\eea
The Ward identity associated with the conservation of the stress tensor (\ref{classicalstresstensor}) is
\bea
\label{identity}
\partial^{\mu} \left\langle T_{\mu\nu} \right\rangle_A =-\left\langle j^{\mu}\right \rangle_A F_{\mu\nu}.
\eea
As the same method as \cite{Bastianelli:2022hmu,Giannotti:2008cv,Armillis:2009pq}, we expand both sides at second order in background gauge field. The right-hand side can be expanded perturbatively as
\bea
&&-\left\langle j^{\mu}\right \rangle_A F_{\mu\nu}=-\left\langle j^{\mu}{\rm{e}}^{i \int d^4y j^{\lambda}(y)A_{\lambda}(y)}\right \rangle F_{\mu\nu}\nonumber \\
&&=-\left\langle j^{\mu}\right \rangle F_{\mu\nu}-i\left\langle j^{\mu} \int d^4y j^{\lambda}(y)A_{\lambda}(y)\right \rangle F_{\mu\nu}+\ldots.
\eea
The conservation of the stress tensor (\ref{identity}) in momentum space at second order in background gauge field is
\bea
\label{wardidentity}
-\frac{i}{2}k_{\rho}T^{\rho\sigma\mu\nu}(p,q)=\Pi^{\rho\mu}(p)(g^{\nu\sigma}q_{\rho}-\delta^{\nu}_{\rho}q^{\sigma})+\Pi^{\rho\nu}(q)(g^{\mu\sigma}p_{\rho}-
\delta^{\mu}_{\rho}p^{\sigma}),
\eea
where the $\Pi^{\mu\nu}(p)$ is the Fourier transform of the two-point function $\left\langle j^{\mu}(x)j^{\nu}(y)\right\rangle$, that is
\bea
\Pi^{\mu\nu}(p)=\int \frac{d^n l}{(2 \pi)^n}\frac{{\rm{tr}}\big[\gamma^{\mu}P_L l\!\!\!/\gamma^{\nu}P_L (l\!\!\!/-p\!\!\!/) \big]}{l^2(l-p)^2}.
\eea
Due to the expression (\ref{expansion2}), the left-hand side of (\ref{wardidentity}) is
\bea
\label{Weydirac1}
&&-\frac{i}{2}k_{\rho}T^{\rho\sigma\mu\nu}(p,q)=\nonumber\\
&&-\frac{1}{8}\int \frac{d^n l}{(2 \pi)^n}\frac{{\rm{tr}}\big[\big(k_{\rho}(2l-p-q)^{\rho}\gamma^{\sigma}+(2l-p-q)^{\sigma}k\!\!\!/ \big)P_L l\!\!\!/\gamma^{\mu}P_L (l\!\!\!/-p\!\!\!/) \gamma^{\nu}P_L (l\!\!\!/-p\!\!\!/-q\!\!\!/)\big]}{l^2(l-p)^2(l-p-q)^2}\nonumber\\
&&+\frac{1}{4}\int \frac{d^n l}{(2 \pi)^n}\frac{{\rm{tr}}\big[\big(k\!\!\!/ g^{\nu\sigma}+\gamma^{\sigma}k^{\nu} \big)P_L l\!\!\!/\gamma^{\mu}P_L (l\!\!\!/-p\!\!\!/) \big]}{l^2(l-p)^2}
+\bigg((p,\mu) \leftrightarrow (q,\nu)\bigg)\nonumber\\
&&=-\frac{1}{8}\int \frac{d^n l}{(2 \pi)^n}\frac{{\rm{tr}}\big[\big(k_{\rho}(2l-p-q)^{\rho}\gamma^{\sigma}+(2l-p-q)^{\sigma}k\!\!\!/ \big) l\!\!\!/\gamma^{\mu}(l\!\!\!/-p\!\!\!/) \gamma^{\nu}\dot{P}_L (l\!\!\!/-p\!\!\!/-q\!\!\!/)\big]}{l^2(l-p)^2(l-p-q)^2}\nonumber\\
&&+\frac{1}{4}\int \frac{d^n l}{(2 \pi)^n}\frac{{\rm{tr}}\big[\big(k\!\!\!/ g^{\nu\sigma}+\gamma^{\sigma}k^{\nu} \big) l\!\!\!/\gamma^{\mu}\dot{P}_L (l\!\!\!/-p\!\!\!/) \big]}{l^2(l-p)^2}
+\bigg((p,\mu) \leftrightarrow (q,\nu)\bigg)\nonumber\\
&&=-\frac{1}{16}\int \frac{d^n l}{(2 \pi)^n}\frac{{\rm{tr}}\big[\big(k_{\rho}(2l-p-q)^{\rho}\gamma^{\sigma}+(2l-p-q)^{\sigma}k\!\!\!/ \big) l\!\!\!/\gamma^{\mu} (l\!\!\!/-p\!\!\!/) \gamma^{\nu} (l\!\!\!/-p\!\!\!/-q\!\!\!/)\big]}{l^2(l-p)^2(l-p-q)^2}\nonumber\\
&&+\frac{1}{8}\int \frac{d^n l}{(2 \pi)^n}\frac{{\rm{tr}}\big[\big(k\!\!\!/ g^{\nu\sigma}+\gamma^{\sigma}k^{\nu} \big) l\!\!\!/\gamma^{\mu} (l\!\!\!/-p\!\!\!/) \big]}{l^2(l-p)^2}
+\bigg((p,\mu) \leftrightarrow (q,\nu)\bigg).
\eea
From (\ref{Weydirac}) and (\ref{Weydirac1}), the Ward identity (\ref{wardidentity}) is equivalent to half the one of a Dirac fermion which has been verified in works \cite{Giannotti:2008cv,Armillis:2009pq}.

In ordinary BMHV regularization, the (\ref{wardidentity}) is not satisfied. On the other hand, the $\Pi^{\mu\nu}(p)$ is
\bea
\Pi^{\mu\nu}(p)&=&\int \frac{d^n l}{(2 \pi)^n}\frac{{\rm{tr}}\big[\bar{\gamma}^{\mu}\dot{P}_L l\!\!\!/\dot{P}_R\bar{\gamma}^{\nu}\dot{P}_L (l\!\!\!/-p\!\!\!/)\dot{P}_R \big]}{l^2(l-p)^2}\nonumber\\
&=&\int \frac{d^n l}{(2 \pi)^n}\frac{{\rm{tr}}\big[\bar{\gamma}^{\mu}\dot{P}_L l\!\!\!/\bar{\gamma}^{\nu}\dot{P}_L (l\!\!\!/-p\!\!\!/) \big]}{l^2(l-p)^2}\nonumber\\
&=&-\frac{i \pi^2 B_0(p^2,0,0)\left(2(n-2)\bar{p}^{\mu}\bar{p}^{\nu}-np^2\bar{g}^{\mu\nu})\right)}{2(n-1)}\nonumber\\
&=&\frac{1}{2}\int \frac{d^n l}{(2 \pi)^n}\frac{{\rm{tr}}\big[\gamma^{\mu}l\!\!\!/\gamma^{\nu} (l\!\!\!/-p\!\!\!/) \big]}{l^2(l-p)^2}+\frac{ip^2\bar{g}^{\mu\nu}}{48\pi^2}\nonumber\\
&\neq& \frac{1}{2}\int \frac{d^n l}{(2 \pi)^n}\frac{{\rm{tr}}\big[\gamma^{\mu}l\!\!\!/\gamma^{\nu} (l\!\!\!/-p\!\!\!/) \big]}{l^2(l-p)^2},
\eea
where the function $B_0(p^2,0,0)$ is defined in (\ref{B000}). So the $\Pi^{\mu\nu}(p)$ in ordinary BMHV regularization does not be transverse.

\section{Fujikawa's method}

In this section, we will study the trace anomaly by the method of Fujikawa \cite{Fujikawa:1979ay,Fujikawa:1980vr}. Following the works of \cite{Diaz:1989nx,Bastianelli:2016nuf,Bastianelli:2018osv,Bastianelli:2019fot,Liu:2022jxz}, we use the Pauli-Villars (PV) regularization \cite{Pauli:1949zm} to compute one-loop anomalies due to the non-invariance of the PV mass term. We use this method to calculate the trace anomaly for a Weyl fermion coupled to an abelian gauge field and a gravity separately.

\subsection{Weyl fermion coupled to an abelian gauge field}

Usually a chiral fermion cannot have a Dirac mass term. There is a possibility to write a mass term with a Majorana mass term which has been used to study the trace anomaly for a chiral fermion  \cite{Bastianelli:2016nuf,Bastianelli:2018osv,Bastianelli:2019fot}. The mass term violates fermion number conservation and breaks the conformal symmetries. We will use a new Dirac mass term which corresponds to the modified BMHV scheme. The new Dirac mass term is
\bea
\label{massterm}
\mathcal{L}_m=-\overline{\psi_L}m^A\psi_L=\mathcal{L}_m^{\dag}.
\eea
Where the mass parameter $m^A$ satisfies
\bea
\{\gamma^5,m^A\}=0, \quad \left[\gamma^{\mu},m^A\right]=0, \quad {m^A}^{\dag}=m^A.
\eea
With the mass term (\ref{massterm}), the equation of motion for a chiral fermion  would be
\bea
\label{eqationsddd}
-D\!\!\!\!/\psi_L-m^A\psi_L=0.
\eea
This equation does not break Lorentz covariance. The equation (\ref{eqationsddd}) is consistent with the equation of motion for a massive Dirac fermion
\bea
-D\!\!\!\!/\psi-m^A\psi=0.
\eea
Classically the gauge current $j^{\mu}$ and axial vector current $j^{\mu 5}$ in the theory satisfy
\bea
\partial_{\mu}j^{\mu}=0, \quad \partial_{\mu}j^{\mu 5}=0.
\eea
The process for calculating the anomaly is the same as the the modified BMHV scheme which has two steps. The first step is simplifying the expression to one which only has one projector $P_L$ or $P_R$. The next step is letting the mass parameter $m^A$ becomes the ordinary real mass $m$ to obtain the final answer. As the modified BMHV scheme, we have some freedom in deciding which current exhibits the anomaly. The form need to be in accordance with the special features of the problem at hand.

Classically, the Lagrangian of (\ref{Lagrangian}) can be expressed in equivalent forms using the the charge conjugated chiral field $\psi_L^c$ rather than $\overline{\psi_L}$
\bea
\mathcal{L}&=&{\psi_L^c}^T C D\!\!\!\!/(A)\psi_L.
\eea
Where the charge conjugated chiral field $\psi_L^c=C^{-1}\overline{\psi_L}^T$ has the opposite chirality of $\psi_L$.
The hermitian conjugate of the Lagrangian is
\bea
\mathcal{L}^{\dag}=-{\psi_L}^T C D\!\!\!\!/(-A)\psi_L^c.
\eea
The kinetic term would be
\bea
\hat{\mathcal{L}}&=&\mathcal{L}+\mathcal{L}^{\dag}=\frac{1}{2}\left({\psi_L^c}^T C D\!\!\!\!/(A)\psi_L-{\psi_L}^T C D\!\!\!\!/(-A)\psi_L^c\right).
\eea
The mass term (\ref{massterm}) can be rewritten as
\bea
\mathcal{L}_m={\psi_L^c}^Tm^A C\psi_L.
\eea
The full massive action $\overline{\mathcal{L}}$ has the form
\bea
\overline{\mathcal{L}}&=&\hat{\mathcal{L}}+\mathcal{L}_m=\frac{1}{2}\left({\psi_L^c}^T C D\!\!\!\!/(A)\psi_L-{\psi_L}^T C D\!\!\!\!/(-A)\psi_L^c\right)+\left({\psi_L^c}^Tm^A C\psi_L\right)\nonumber\\
&=&\frac{1}{2}\phi^T T\mathcal{O}\phi+\frac{1}{2} \phi^T m^A \widetilde{T} \phi,
\eea
where $\phi$ is a column vector
\bea
\phi=\left(
  \begin{array}{c}
    \psi_L  \\
    C^{-1}\overline{\psi_L}^T  \\
  \end{array}
\right)=
\left(
  \begin{array}{c}
    \psi_L  \\
    \psi_L^c  \\
  \end{array}
\right).
\eea
Here the matrix $T$ and $\mathcal{O}$ are defined as
\bea
\label{TOP}
T=\left(
  \begin{array}{cc}
    C P_L&\ \ 0 \\
    0 &\  \ C P_R \\
  \end{array}
\right),
\quad
\mathcal{O}=\left(
  \begin{array}{cc}
    0 &\ \ -D\!\!\!\!/(-A)P_R \\
    D\!\!\!\!/(A) P_L &\ \ 0 \\
  \end{array}
\right),
\quad
\widetilde{T}=\left(
  \begin{array}{cc}
    0 &\ \ 0 \\
    2C P_L &\  \ 0\\
  \end{array}
\right).
\eea

We introduce the PV field $\theta$ with mass $M^A$ and $\theta_L=P_L\theta$. The Lagrangian of PV field is
\bea
\overline{\mathcal{L}}_{PV}=\frac{1}{2}\chi^T T\mathcal{O}\chi+\frac{1}{2}  \chi^TM^A \widetilde{T} \chi,
\eea
where $\chi$ is a column vector
\bea
\chi=\left(
  \begin{array}{c}
    \theta_L  \\
    C^{-1}\overline{\theta_L}^T  \\
  \end{array}
\right)=
\left(
  \begin{array}{c}
    \theta_L \\
    \theta_L^c  \\
  \end{array}
\right).
\eea
Usually, we need to introduce a set of PV fields to regulate and cancel all possible one-loop divergences. To obtain the anomaly, only one PV field is enough. In the following, we will derive the formula for calculating the anomalies.
The quantum theory is defined by the path integral
\bea
\label{quantumtheory}
Z={\rm{e}}^{iW[e^{\ a}_{\mu}]}=\int \mathcal{D}\phi\mathcal{D}\chi {\rm{e}}^{i\widetilde{S}(\phi,\chi;e^{\ a}_{\mu})}.
\eea
Where the regulated action has the form
\bea
\widetilde{S}(\phi,\chi;e^{\ a}_{\mu})=\int d^4x \left(\hat{\mathcal{L}}+\overline{\mathcal{L}}_{{\rm{PV}}}\right).
\eea
With a dummy change of integration variables in (\ref{quantumtheory}), we get
\bea
\label{pathintegral}
0&=&\int \mathcal{D}\phi'\mathcal{D}\chi' {\rm{e}}^{i\widetilde{S}(\phi',\chi';e^{\ a}_{\mu})}-\int \mathcal{D}\phi\mathcal{D}\chi {\rm{e}}^{i\widetilde{S}(\phi,\chi;e^{\ a}_{\mu})}\nonumber\\
&=&\int \mathcal{D}\phi\mathcal{D}\chi {\rm{e}}^{i\widetilde{S}(\phi',\chi';e^{\ a}_{\mu})}-\int \mathcal{D}\phi\mathcal{D}\chi {\rm{e}}^{i\widetilde{S}(\phi,\chi;e^{\ a}_{\mu})}.
\eea
Here the jacobian of PV field cancels the jacobian of the original fields $\phi$. Defining the infinitesimal change in the field variable as
\bea
\delta \phi=\phi'-\phi, \quad  \delta \chi=\chi'-\chi,
\eea
 the change of the action is
\bea
\widetilde{S}(\phi',\chi';e^{\ a}_{\mu})=\widetilde{S}(\phi,\chi;e^{\ a}_{\mu})+\delta\widetilde{S}(\phi,\chi;e^{\ a}_{\mu}).
\eea
The (\ref{pathintegral}) becomes
\bea
\label{pathintegral1}
0&=&\int \mathcal{D}\phi\mathcal{D}\chi {\rm{e}}^{i\widetilde{S}(\phi,\chi;e^{\ a}_{\mu})+i\delta\widetilde{S}(\phi,\chi;e^{\ a}_{\mu})}-\int \mathcal{D}\phi\mathcal{D}\chi {\rm{e}}^{i\widetilde{S}(\phi,\chi;e^{\ a}_{\mu})}\nonumber\\
&=&\int \mathcal{D}\phi\mathcal{D}\chi {\rm{e}}^{i\widetilde{S}(\phi,\chi;e^{\ a}_{\mu})}\left({\rm{e}}^{i\delta\widetilde{S}(\phi,\chi;e^{\ a}_{\mu})}-1\right)\nonumber\\
&=&\int \mathcal{D}\phi\mathcal{D}\chi {\rm{e}}^{i\widetilde{S}(\phi,\chi;e^{\ a}_{\mu})}\left(i\delta\widetilde{S}(\phi,\chi;e^{\ a}_{\mu})\right)\nonumber\\
&=& \left\langle i\delta\widetilde{S}(\phi,\chi;e^{\ a}_{\mu})\right\rangle.
\eea
Where the $\delta\widetilde{S}(\phi,\chi;e^{\ a}_{\mu})$ is
\bea
&&\delta\widetilde{S}(\phi,\chi;e^{\ a}_{\mu})=\delta\int d^4x \left(\hat{\mathcal{L}}+\overline{\mathcal{L}}_{{\rm{PV}}}\right)\nonumber\\
&&=\delta\int d^4x \left(\frac{1}{2}\phi^T T\mathcal{O}\phi+\frac{1}{2}\chi^T T\mathcal{O}\chi\right)+\delta\int d^4x \left(\frac{1}{2}\chi^T M^A \widetilde{T} \chi\right).
\eea
The expression (\ref{pathintegral1}) can be reformulated as
\bea
\label{pathintegral2}
\left\langle\delta\int d^4x (\frac{1}{2}\phi^T T\mathcal{O}\phi+\frac{1}{2}\chi^T T\mathcal{O}\chi)\right\rangle=-\left\langle\delta\int d^4x (\frac{1}{2}  \chi^T M^A \widetilde{T} \chi)\right\rangle.
\eea
Under infinitesimal gauge transformation (\ref{u1}), the left-hand side becomes
\bea
\left\langle\delta\int d^4x (\frac{1}{2}\phi^T T\mathcal{O}\phi+\frac{1}{2}\chi^T T\mathcal{O}\chi)\right\rangle&=&\int d^4x \alpha(x)\partial_{\mu}\left\langle i\overline{\psi_L}\gamma^{\mu}\psi_L+i\overline{\theta_L}\gamma^{\mu}\theta_L\right\rangle \nonumber\\
&=&\int d^4x \alpha(x)\partial_{\mu}\left\langle j^{\mu,reg}\right\rangle.
\eea
Here we define the regulated gauge current $j^{\mu,reg}$, which is
\bea
j^{\mu,reg}=i\overline{\psi_L}\gamma^{\mu}\psi_L+i\overline{\theta_L}\gamma^{\mu}\theta_L.
\eea
Then the (\ref{pathintegral2}) becomes
\bea
\int d^4x \alpha(x)\partial_{\mu}\left\langle j^{\mu,reg}\right\rangle=-\left\langle\delta\int d^4x (\frac{1}{2} \chi^T M^A \widetilde{T} \chi)\right\rangle.
\eea
 From the gauge transformation (\ref{u1}), we find
\bea
\partial_{\mu}\left\langle j^{\mu,reg}\right\rangle=0.
\eea
The gauge current $j^{\mu}$ which couples to external gauge field $A_{\mu}$ is conserved.

We consider a local chiral transformation, which is
\bea
\label{chiral}
\begin{cases}
   \psi_L(x) \quad \rightarrow \quad \psi_L'(x)={\rm{e}}^{i \beta(x)\gamma^5}\psi_L(x)\\
    \overline{\psi_L}(x)\quad \rightarrow \quad \overline{\psi_L}'(x)=\overline{\psi_L}(x){\rm{e}}^{i \beta(x)\gamma^5}\\
    A_{\mu}(x) \quad \rightarrow \quad A_{\mu}'(x)=A_{\mu}(x)\\
    e^{\ a}_{\mu}(x)\quad \rightarrow \quad e'^{\ a}_{\mu}(x)=e^{\ a}_{\mu}(x).
 \end{cases}
\eea
Under the infinitesimal local chiral transformation (\ref{chiral}), the (\ref{pathintegral2}) becomes
\bea
&&\int d^4x \beta(x)\partial_{\mu}\left\langle j^{\mu5,reg}\right\rangle=-\left\langle\delta\int d^4x (\frac{1}{2} \chi^T M^A \widetilde{T} \chi)\right\rangle\nonumber\\
&&=\left\langle \delta \int d^4x  \frac{1}{2}\left( \overline{\theta}P_R M^A \theta+\overline{\theta} M^A P_L\theta\right)\right\rangle.
\eea
As the same as the modified BMHV regularization, there are different ways to express the results with $M^A$ which can lead to different results when $M^A$ becomes the real mass $M$. To obtain the correct chiral anomaly for axial vector current, the expression is
\bea
&&\int d^4x \beta(x)\partial_{\mu}\left\langle j^{\mu5,reg}\right\rangle=\left\langle \delta \int d^4x  \frac{1}{2}\left( \overline{\theta}P_R M \theta+\overline{\theta} M P_L\theta\right)\right \rangle\nonumber\\
&&=\left\langle \delta \int d^4x  \frac{1}{2}\left( \overline{\theta} M \theta\right)\right\rangle= \int d^4x i \beta(x)\left\langle \left( \overline{\theta} M \gamma^5 \theta\right)\right\rangle.
\eea
Using PV propagator
\bea
\label{PVpropagator}
\left\langle  \theta  \overline{\theta}\right\rangle=\frac{i}{D\!\!\!\!/+M},
\eea
we have
\bea
&&\int d^4x \beta(x)\partial_{\mu}\left\langle j^{\mu5,reg}\right\rangle=-\lim_{M\rightarrow \infty} {\rm{Tr}}\left[\beta(x)\gamma^5 M\frac{1}{D\!\!\!\!/+M}\right]\nonumber\\
&&=-\lim_{M\rightarrow \infty} {\rm{Tr}}\left[\beta(x)\gamma^5 {\rm{e}}^{\frac{D\!\!\!\!/^2}{M^2}}\right].
\eea
With the standard Fujikawa's method (e.g. \cite{Weinberg:1996kr,Bilal:2008qx}), the axial vector current satisfies
\bea
\partial_{\mu}\left\langle j^{\mu5,reg}\right\rangle=\frac{1}{32\pi^2}\epsilon^{\mu\nu\rho\sigma}F_{\mu\nu}(x)F_{\rho\sigma}(x).
\eea

To calculate the trace anomaly, we need to consider the local Weyl symmetry (\ref{weyl}). From the (\ref{quantumtheory}), we have
\bea
\label{deltasigma}
\delta_{\sigma(x)}W[e^{\ a}_{\mu}]&=&-i\delta_{\sigma(x)}\ln Z[e^{\ a}_{\mu}]=\left\langle \delta_{\sigma(x)}\widetilde{S}(\phi,\chi;e^{\ a}_{\mu})\right\rangle
=-\int d^4x e \sigma(x)\left\langle T^{\mu}_{ \  \mu}(x)\right\rangle \nonumber\\
&=&-i\frac{1}{Z[e^{\ a}_{\mu}]}\delta_{\sigma(x)}Z[e^{\ a}_{\mu}]=-i\frac{1}{Z[e^{\ a}_{\mu}]}\left[\int \mathcal{D}\phi\mathcal{D}\chi {\rm{e}}^{i\widetilde{S}(\phi,\chi;e'^{\ a}_{\mu})}
-\int \mathcal{D}\phi\mathcal{D}\chi {\rm{e}}^{i\widetilde{S}(\phi,\chi;e^{\ a}_{\mu})}\right]\nonumber\\
&=&-i\frac{1}{Z[e^{\ a}_{\mu}]}\left[\int \mathcal{D}\phi\mathcal{D}\chi {\rm{e}}^{i\widetilde{S}(\phi',\chi';e'^{\ a}_{\mu})}
-\int \mathcal{D}\phi\mathcal{D}\chi {\rm{e}}^{i\widetilde{S}(\phi,\chi;e^{\ a}_{\mu})}\right]\nonumber\\
&=&\left\langle \delta\int d^4x \left(\frac{1}{2}e \chi^T M^A \widetilde{T} \chi \right)\right\rangle=
\left\langle \int d^4x \left(\frac{1}{2}e \sigma(x)\chi^T M^A \widetilde{T} \chi \right)\right\rangle\nonumber\\
 &=&-\int d^4x e \sigma(x)\left\langle \overline{\theta_L} M^A \theta_L \right\rangle=-\int d^4x e \sigma(x)\left\langle \overline{\theta} M^A P_L\theta \right\rangle\nonumber\\
 &=&-\int d^4x e \sigma(x)\left\langle \overline{\theta} M P_L\theta \right\rangle.
\eea
Using PV propagator (\ref{PVpropagator}), we obtain
\bea
&&\int d^4x e \sigma(x)\left\langle T^{\mu}_{ \  \mu}(x)\right\rangle=i\lim_{M\rightarrow \infty} {\rm{Tr}}\left[\sigma(x)P_LM\frac{1}{D\!\!\!\!/+M}\right]\nonumber\\
&&=i\lim_{M\rightarrow \infty} {\rm{Tr}}\left[\sigma(x)P_L\left(1+\frac{D\!\!\!\!/}{M}\right)^{-1}\right]=i\lim_{M\rightarrow \infty} {\rm{Tr}}\left[\sigma(x)P_L{\rm{e}}^{\frac{D\!\!\!\!/^2}{M^2}}\right].
\eea
The final result can be obtained by the heart kernel method (the formula (\ref{a4})), i.e.
\bea
\left\langle T^{\mu}_{ \  \mu}(x)\right\rangle&=&\frac{1}{360}\frac{1}{(4\pi)^2}{\rm{tr}}\{90E^2\gamma^5 +90E^2 +15\Omega_{\mu\nu}\Omega^{\mu\nu}\}\nonumber\\
&=&-\frac{i}{64\pi^2}\epsilon^{\mu\nu\rho\sigma}F_{\mu\nu}(x)F_{\rho\sigma}(x)-\frac{1}{48\pi^2}F_{\mu\nu}(x)F^{\mu\nu}(x).
\eea
Where the $E$ and $\Omega_{\mu\nu}$  are
\bea
E=\frac{1}{4}\left[\gamma^{\mu},\gamma^{\nu}\right]F_{\mu\nu}, \quad \Omega_{\mu\nu}=F_{\mu\nu}.
\eea
There is an extra $i$ which comes from Euclidean space back to Minkowski space-time. The result agree with the previous result (\ref{traceanomaly69}).

There is another result for the expression (\ref{deltasigma}), that is
\bea
-\int d^4x e \sigma(x)\left\langle T^{\mu}_{ \  \mu}(x)\right\rangle&=&-\frac{1}{2}\int d^4x e \sigma(x)\left\langle \overline{\theta}(P_R M^A+ M^A P_L)\theta \right\rangle\nonumber\\
 &=&-\frac{1}{2}\int d^4x e \sigma(x)\left\langle \overline{\theta} M\theta \right\rangle.
\eea
The result only have parity-even term and is not the same as our previous one, so we omit it.
\subsection{Trace anomaly of a Weyl fermion coupled to gravity}
We now study the trace anomaly of a Weyl fermion coupled to gravity by the new method. The Lagrangian is
\bea
\label{Lagrangianweylgravity}
\mathcal{L}=-e\overline{\psi_L}\gamma^{\mu}\check{\nabla}_{\mu}\psi_L,
\eea
where the covariant derivative $\check{\nabla}_{\mu}$ is
\bea
\check{\nabla}_{\mu}=\partial_{\mu}+\frac{1}{4}\omega_{\mu a b}\gamma^a \gamma^b.
\eea
Using the same method, the trace of energy-momentum tensor  $\check{T}^{\mu}_{ \  \mu}$ satisfies
\bea
\int d^4x e \sigma(x)\left\langle \check{T}^{\mu}_{ \  \mu}(x)\right\rangle=i\lim_{M\rightarrow \infty} {\rm{Tr}}\left[\sigma(x)P_L{\rm{e}}^{\frac{\check{\nabla}\!\!\!\!/^2}{M^2}}\right].
\eea
From this, we find that the $\left\langle\check{T}^{\mu}_{ \  \mu}\right\rangle$ has the parity-odd term $\left\langle \check{T}^{\mu}_{ \  \mu}(x)\right\rangle^{odd}$, i.e.
\bea
\int d^4x e \sigma(x)\left\langle \check{T}^{\mu}_{ \  \mu}(x)\right\rangle^{odd}&=&i\lim_{M\rightarrow \infty} {\rm{Tr}}\left[\frac{1}{2}\sigma(x)\gamma^5{\rm{e}}^{\frac{\check{\nabla}\!\!\!\!/^2}{M^2}}\right]\nonumber\\
&=&\frac{i}{1536\pi^2}\int d^4x \sigma(x) e R_{\sigma\rho\mu\nu}R^{\mu\nu}_{\ \ ij}\epsilon^{\sigma\rho ij}.
\eea
The calculation is the same as our previous work \cite{Liu:2022jxz}. We once again obtain the parity-odd term of trace anomaly for a Weyl fermion coupled to gravity \cite{Bonora:2014qla,Liu:2022jxz}.
\section{Conclusions and Discussions }

In this paper, we have studied the problem of trace anomaly for chiral fermions. Though the BMHV regularization give mathematically consistent results to arbitrary loop orders for non-chiral theories, it have some obstacles to apply this scheme to chiral theories. We used a modified BMHV regularization to calculate the trace anomaly for a Weyl fermion coupled to an abelian gauge field. To confirm these results, we also used the Fujikawa's method with a new Dirac mass term for Pauli-Villars field to calculate it. In the calculation for chiral anomaly, there are some freedom in deciding which current exhibits the anomaly.  In our new approach, the evaluation of the anomaly depends on the choice of the $P_L$ or $P_R$ place in amplitudes. The result is required to be in accordance with the special features of the problem at hand. This is the one property of the anomaly, because the anomaly can move from one current to others by the addition of a local counterterm to the classical action. The chiral anomaly are given by non-trivial BRST cohomology classes at ghost number one on the space of local functionals. We show once again that the trace anomaly of energy-momentum tensor for a chiral fermion has parity-odd term.

In my opinion, we have not yet understand the trace anomaly for a chiral fermion completely. There is still quite a lot of work to do on this problem in the future.


\section*{Acknowledgements}
This work is supported by Chinese Universities Scientific Fund Grant No. 2452018158.

\appendix
\section{Conventions and notations}
Our Minkowski space signature is $(-+++)$. The gamma matrices satisfy
\bea
\{\gamma^{\mu},\gamma^{\nu}\}=2 g^{\mu\nu}
\eea
as well as $(i \gamma^0)^{\dag}=i \gamma^0$, $(\gamma^j)^{\dag}=\gamma^j$ and $\gamma_0=-\gamma^0$ in four dimensional space-time. In terms of $2\times 2$ blocks they are given by
\bea
\label{dirac}
\gamma^0=-i\left(
  \begin{array}{cc}
    0&\ \ 1_{2\times 2} \\
    1_{2\times 2}&\  \ 0 \\
  \end{array}
\right),
\quad
\gamma^i=-i\left(
  \begin{array}{cc}
    0 &\ \ \sigma^i \\
    -\sigma^i &\ \ 0 \\
  \end{array}
\right),
\eea
where $\sigma^i$ are the Pauli matrices. The gamma matrices also have the relation
\bea
{\gamma^{\mu}}^{\dagger}=-(i \gamma^0)\gamma^{\mu}(i \gamma^0).
\eea
We define $\bar{\psi}=\psi^{\dagger}i\gamma^0$ and the chiral matrix $\gamma^5$
\bea
\gamma^5=-i\gamma^0\gamma^1\gamma^2\gamma^3=\frac{i}{4!}\epsilon^{\mu\nu\rho\sigma}\gamma_{\mu}\gamma_{\nu}\gamma_{\rho}\gamma_{\sigma}=\left(
  \begin{array}{cc}
    1_{2\times 2} &\ \ 0 \\
    0 &\ \ -1_{2\times 2} \\
  \end{array}
  \right).
\eea
The completely antisymmetric tensor $\epsilon^{\mu\nu\rho\sigma}$ is defined as
\bea
\epsilon^{0123}=+1, \quad \epsilon_{0123}=-1,
\eea
so that we have
\bea
\label{trgamma}
{\rm{tr}}\left(\gamma^5\gamma^{\mu}\gamma^{\nu}\gamma^{\rho}\gamma^{\sigma}\right)=4i\epsilon^{\mu\nu\rho\sigma}.
\eea
The hermitian chiral matrix $\gamma^5$ is used to define the chiral projectors
\bea
P_L=\frac{1+\gamma^5}{2}, \quad P_R=\frac{1-\gamma^5}{2}.
\eea
The charge conjugation of a field $\psi$ is
\bea
\psi^c=C^{-1}\bar{\psi}^T,
\eea
where the charge conjugation matrix $C$ satisfies
\bea
C\gamma^{\mu}C^{-1}=-\gamma^{\mu T}.
\eea
In chiral representation the charge conjugation matrix $C$ may be given by
\bea
C=i\gamma^0\gamma^2=-i \left(
  \begin{array}{cc}
    \sigma^2 &\ \ 0 \\
    0 &\ \ -\sigma^2 \\
  \end{array}
\right)
\eea
and satisfies
\bea
C=-C^T=-C^{-1}=-C^{\dagger}=C^{\ast}.
\eea

There are many proposals for the definition of chiral $\gamma^5$ matrix in dimensional regularization. We list two prescriptions as follows  \cite{Jegerlehner:2000dz}:
\begin{enumerate}
\item Naive dimensional regularization

In naive dimensional regularization, the algebra of $\gamma$ matrices satisfy
\bea
\{\gamma^{\mu},\gamma^{\nu}\}=2 g^{\mu\nu}, \quad \{\gamma^{\mu},\gamma^{5}\}=0
\eea
for dimensions of space-time $n$. The cyclicity of the trace and associativity of the algebra of $\gamma$ matrices follow that \cite{Baikov:1991qz,Novotny:1994yx}
\bea
n(n-2)(n-4){\rm{tr}}\left(\gamma^5\gamma^{\mu}\gamma^{\nu}\gamma^{\rho}\gamma^{\sigma}\right)=0.
\eea
If ${\rm{tr}}\left(\gamma^5\gamma^{\mu}\gamma^{\nu}\gamma^{\rho}\gamma^{\sigma}\right)$ is regarded as analytic function of dimension $n$, then it must also be equal to zero at $n=4$, and this contradicts with (\ref{trgamma}). There need a modified naive dimensional regularization.
\item Breitenlohner-Maison-'t Hooft-Veltman regularization

In Breitenlohner-Maison-'t Hooft-Veltman (BMHV) regularization, the usual $n$-dimensional objects (denoted by a dot to distinguished from the naive dimensional regularization) decompose into a four-dimensional part (denoted by a bar) and an $(n-4)$-dimensional part (denoted by a hat):
\bea
\dot{g}^{\mu\nu}=\bar{g}^{\mu\nu}+\hat{g}^{\mu\nu},\quad \dot{\gamma}^{\mu}=\bar{\gamma}^{\mu}+\hat{\gamma}^{\mu},\quad \dot{p}^{\mu}=\bar{p}^{\mu}+\hat{p}^{\mu},\quad \cdots.
\eea
In this regularization, 't Hooft-Veltman \cite{tHooft:1972tcz} using the definition
\bea
\dot{\gamma}^5=-i\bar{\gamma}^0\bar{\gamma}^1\bar{\gamma}^2\bar{\gamma}^3
\eea
in $n$ dimensions. This definition has consequence that
\bea
\begin{cases}
   \{\dot{\gamma}^5, \bar{\gamma}^{\mu}\}=0 & {\rm{for}}\:\: \mu=0,1,2,3; \\
    [\dot{\gamma}^5, \hat{\gamma}^{\mu}]=0 &{\rm{for}}\:\: \mu=4,\ldots,n-1.
 \end{cases}
\eea
To distinguish the naive dimensional regularization, two chiral projectors in BMHV regularization are denoted by
\bea
\dot{P}_L=\frac{1+\dot{\gamma}^5}{2}, \quad \dot{P}_R=\frac{1-\dot{\gamma}^5}{2}.
\eea

\end{enumerate}

\section{The heat kernel method}
In this section, we present some definitions and useful formulae from the review paper \cite{Vassilevich:2003xt}. Let $M$ be a smooth compact Riemannian manifold of dimension $n$ with metric tensor $g_{\mu\nu}$ and spin connection $\omega_{\mu}^{ab}$. The spin connection is
\bea
\omega_{\mu ab}=e_{\nu a}(\partial_{\mu} e^{\nu}_{b}+e^{\sigma}_{b} \Gamma_{\sigma\ \mu}^{\ \nu}),
\eea
where the Levi-Civita connection $\Gamma_{\sigma\ \mu}^{\ \nu}$ is defined by
\bea
\Gamma_{\sigma\ \mu}^{\ \nu}=\frac{1}{2}g^{\nu \rho}(\partial_{\sigma}g_{\mu \rho}+\partial_{\mu}g_{\sigma \rho}-\partial_{\rho}g_{\sigma \mu}).
\eea
The field strength of the connection $\omega$ and the Riemann curvature tensor $R^{\mu}_{\ \nu\rho\sigma}$ are
\bea
\Omega_{\mu\nu}&=&\partial_{\mu}\omega_{\nu}-\partial_{\nu}\omega_{\mu}+\omega_{\mu}\omega_{\nu}-\omega_{\nu}\omega_{\mu}, \nonumber\\
R^{\mu}_{\ \nu\rho\sigma}&=&\partial_{\sigma}\Gamma^{\mu}_{\nu\rho}-\partial_{\rho}\Gamma^{\mu}_{\nu\sigma}+\Gamma^{\lambda}_{\nu\rho}
\Gamma^{\mu}_{\lambda\sigma}-\Gamma^{\lambda}_{\nu\sigma}\Gamma^{\mu}_{\lambda\rho}.
\eea
Suppose the $D$ be self-adjoint operator and $f$ be an
auxiliary smooth function on $M$, there is an asymptotic
expansion as $t\rightarrow 0$
\bea
\label{trdd}
{\rm{Tr}}_{L^2}\left(f{\rm{e}}^{-tD}\right)\cong \sum_{k\geq 0}t^{\frac{(k-n)}{2}}a_k(f,D).
\eea
The leading heat kernel coefficients $a_4(f,D)$ is known as \cite{DeWitt:1964mxt,McKean:1967xf}
\bea
\label{a4}
a_4(f,D)&=&\frac{1}{360}\frac{1}{(4\pi)^{\frac{n}{2}}}\int_M d^nx \sqrt{g}{\rm{tr}}_V\{f(60E_{;k}^{\ \ k}+60RE+180E^2 \nonumber\\
&&+12R_{;k}^{\ \ k}
+5R^2-2R_{ij}R^{ij}+2R_{ijkl}R^{ijkl}+30\Omega_{ij}\Omega^{ij})\}.
\eea
Where the $;$ denotes multiple covariant differentiation with respect to the Levi-Civita connection of $M$. The $R_{\mu\nu}:=
R^{\sigma}_{\ \mu\nu\sigma}$ is the Ricci tensor and $R:=R^{\mu}_{\mu}$ is the scalar curvature.
We consider the operator $D\!\!\!\!/$ to be the form
\bea
D\!\!\!\!/=i\gamma^{\mu}(\partial_{\mu}+A_{\mu}+\frac{1}{8}[\gamma_a,\gamma_b]\omega^{ab}_{\mu})
\eea
The operator $D$ in (\ref{trdd}) is $D=D\!\!\!\!/^2$. The $E$ and $\Omega_{\mu\nu}$ in (\ref{a4}) associated with the operator $D\!\!\!\!/$ are
\bea
E=\frac{1}{4}\left[\gamma^{\mu},\gamma^{\nu}\right]F_{\mu\nu}
-\frac{1}{4}R, \quad \Omega_{\mu\nu}=F_{\mu\nu}-\frac{1}{4}\gamma^{a}\gamma^{b}R_{ab\mu\nu}.
\eea



\begin{thebibliography}{99}

\bibitem{Adler:1969gk}
S.~L.~Adler,
Phys. Rev. \textbf{177} (1969), 2426-2438
doi:10.1103/PhysRev.177.2426.
\bibitem{Bell:1969ts}
J.~S.~Bell and R.~Jackiw,
Nuovo Cim. A \textbf{60} (1969), 47-61
doi:10.1007/BF02823296.
\bibitem{Gross:1972pv}
D.~J.~Gross and R.~Jackiw,
Phys. Rev. D \textbf{6} (1972), 477-493
doi:10.1103/PhysRevD.6.477.
\bibitem{Liu:2021pmt}
C.~Y.~Liu,
Mod. Phys. Lett. A \textbf{36} (2021) no.12, 2150080
doi:10.1142/S0217732321500802
\bibitem{Liu:2022zdo}
C.~Y.~Liu,
Mod. Phys. Lett. A \textbf{37} (2022) no.21, 2250137
doi:10.1142/S0217732322501371
\bibitem{Alvarez-Gaume:1983ihn}
L.~Alvarez-Gaume and E.~Witten,
Nucl. Phys. B \textbf{234} (1984), 269
doi:10.1016/0550-3213(84)90066-X
\bibitem{Capper:1974ic}
D.~M.~Capper and M.~J.~Duff,
Nuovo Cim. A \textbf{23} (1974), 173-183
doi:10.1007/BF02748300
\bibitem{Capper:1975ig}
D.~M.~Capper and M.~J.~Duff,
Phys. Lett. A \textbf{53} (1975), 361
doi:10.1016/0375-9601(75)90030-4
\bibitem{Deser:1976yx}
S.~Deser, M.~J.~Duff and C.~J.~Isham,
Nucl. Phys. B \textbf{111} (1976), 45-55
doi:10.1016/0550-3213(76)90480-6
\bibitem{Christensen:1976vb}
S.~M.~Christensen,
Phys. Rev. D \textbf{14} (1976), 2490-2501
doi:10.1103/PhysRevD.14.2490
\bibitem{Duff:1977ay}
M.~J.~Duff,
Nucl. Phys. B \textbf{125} (1977), 334-348
doi:10.1016/0550-3213(77)90410-2
\bibitem{Dowker:1976zf}
J.~S.~Dowker and R.~Critchley,
Phys. Rev. D \textbf{16} (1977), 3390
doi:10.1103/PhysRevD.16.3390
\bibitem{Wald:1978pj}
R.~M.~Wald,
Phys. Rev. D \textbf{17} (1978), 1477-1484
doi:10.1103/PhysRevD.17.1477
\bibitem{Bonora:1985cq}
L.~Bonora, P.~Pasti and M.~Bregola,
Class. Quant. Grav. \textbf{3} (1986), 635
doi:10.1088/0264-9381/3/4/018
\bibitem{Bonora:1984ic}
L.~Bonora, P.~Pasti and M.~Tonin,
J. Math. Phys. \textbf{27} (1986), 2259
doi:10.1063/1.526998
\bibitem{Duff:1993wm}
M.~J.~Duff,
Class. Quant. Grav. \textbf{11} (1994), 1387-1404
doi:10.1088/0264-9381/11/6/004
[arXiv:hep-th/9308075 [hep-th]].

\bibitem{Nakayama:2012gu}
Y.~Nakayama,
Nucl. Phys. B \textbf{859} (2012), 288-298
doi:10.1016/j.nuclphysb.2012.02.006
[arXiv:1201.3428 [hep-th]].

\bibitem{Bonora:2014qla}
L.~Bonora, S.~Giaccari and B.~Lima de Souza,
JHEP \textbf{07}, 117 (2014)
doi:10.1007/JHEP07(2014)117
[arXiv:1403.2606 [hep-th]].
\bibitem{Bonora:2015nqa}
L.~Bonora, A.~D.~Pereira and B.~Lima de Souza,
JHEP \textbf{06} (2015), 024
doi:10.1007/JHEP06(2015)024
[arXiv:1503.03326 [hep-th]].


\bibitem{Bonora:2017gzz}
L.~Bonora, M.~Cvitan, P.~Dominis Prester, A.~Duarte Pereira, S.~Giaccari and T.~\v{S}temberga,
Eur. Phys. J. C \textbf{77} (2017) no.8, 511
doi:10.1140/epjc/s10052-017-5071-7
[arXiv:1703.10473 [hep-th]].


\bibitem{Bonora:2018obr}
L.~Bonora, M.~Cvitan, P.~Dominis Prester, S.~Giaccari, M.~Pauli\v{s}i\'c and T.~\v{S}temberga,
Eur. Phys. J. C \textbf{78} (2018) no.8, 652
doi:10.1140/epjc/s10052-018-6141-1
[arXiv:1807.01249 [hep-th]].

\bibitem{Bonora:2020upc}
L.~Bonora, R.~Soldati and S.~Zalel,
Universe \textbf{6} (2020) no.8, 111
doi:10.3390/universe6080111
[arXiv:2006.04546 [hep-th]].


\bibitem{Liu:2022jxz}
C.~Y.~Liu,
Nucl. Phys. B \textbf{980} (2022), 115840
doi:10.1016/j.nuclphysb.2022.115840
[arXiv:2202.13893 [hep-th]].

\bibitem{Bastianelli:2016nuf}
F.~Bastianelli and R.~Martelli,
JHEP \textbf{11} (2016), 178
doi:10.1007/JHEP11(2016)178
[arXiv:1610.02304 [hep-th]].
\bibitem{Bastianelli:2018osv}
F.~Bastianelli and M.~Broccoli,
Eur. Phys. J. C \textbf{79} (2019) no.4, 292
doi:10.1140/epjc/s10052-019-6799-z
[arXiv:1808.03489 [hep-th]].


\bibitem{Bastianelli:2019fot}
F.~Bastianelli and M.~Broccoli,
JHEP \textbf{10} (2019), 241
doi:10.1007/JHEP10(2019)241
[arXiv:1908.03750 [hep-th]].
\bibitem{Bastianelli:2019zrq}
F.~Bastianelli and M.~Broccoli,
Eur. Phys. J. C \textbf{80} (2020) no.3, 276
doi:10.1140/epjc/s10052-020-7782-4
[arXiv:1911.02271 [hep-th]].


\bibitem{Bastianelli:2022hmu}
F.~Bastianelli and L.~Chiese,
Nucl. Phys. B \textbf{983} (2022), 115914
doi:10.1016/j.nuclphysb.2022.115914
[arXiv:2203.11668 [hep-th]].

\bibitem{Frob:2019dgf}
M.~B.~Fr\"ob and J.~Zahn,
JHEP \textbf{10} (2019), 223
doi:10.1007/JHEP10(2019)223
[arXiv:1904.10982 [hep-th]].

\bibitem{Abdallah:2021eii}
S.~Abdallah, S.~A.~Franchino-Vi\~nas and M.~B.~Fr\"ob,
JHEP \textbf{03} (2021), 271
doi:10.1007/JHEP03(2021)271
[arXiv:2101.11382 [hep-th]].
\bibitem{Bonora:2022izj}
L.~Bonora,
EPL \textbf{139} (2022) no.4, 44001
doi:10.1209/0295-5075/ac83e9
[arXiv:2207.03279 [hep-th]].



\bibitem{tHooft:1972tcz}
G.~'t Hooft and M.~J.~G.~Veltman,
Nucl. Phys. B \textbf{44}, 189-213 (1972)
doi:10.1016/0550-3213(72)90279-9
\bibitem{Breitenlohner:1977hr}
P.~Breitenlohner and D.~Maison,
Commun. Math. Phys. \textbf{52}, 11-38 (1977)
doi:10.1007/BF01609069

\bibitem{Breitenlohner:1975hg}
P.~Breitenlohner and D.~Maison,
Commun. Math. Phys. \textbf{52}, 39 (1977)
doi:10.1007/BF01609070

\bibitem{Breitenlohner:1976te}
P.~Breitenlohner and D.~Maison,
Commun. Math. Phys. \textbf{52}, 55 (1977)
doi:10.1007/BF01609071

\bibitem{Rufa:1990hg}
G.~Rufa,
Annalen Phys. \textbf{47} (1990), 6-26

\bibitem{Fujikawa:1979ay}
K.~Fujikawa,
Phys. Rev. Lett. \textbf{42} (1979), 1195-1198
doi:10.1103/PhysRevLett.42.1195
\bibitem{Fujikawa:1980vr}
K.~Fujikawa,
Phys. Rev. Lett. \textbf{44} (1980), 1733
doi:10.1103/PhysRevLett.44.1733

\bibitem{Pauli:1949zm}
W.~Pauli and F.~Villars,
Rev. Mod. Phys. \textbf{21} (1949), 434-444
doi:10.1103/RevModPhys.21.434

\bibitem{Jegerlehner:2000dz}
F.~Jegerlehner,
Eur. Phys. J. C \textbf{18} (2001), 673-679
doi:10.1007/s100520100573
[arXiv:hep-th/0005255 [hep-th]].
\bibitem{Weinberg:1996kr}
S.~Weinberg,
``The quantum theory of fields. Vol. 2: Modern applications,''
Cambridge University Press (2013).
\bibitem{Bilal:2008qx}
A.~Bilal,
[arXiv:0802.0634 [hep-th]].
\bibitem{Peskin:1995ev}
M.~E.~Peskin and D.~V.~Schroeder,
``An Introduction to quantum field theory,''
 Westview Press (1995).
\bibitem{Giannotti:2008cv}
M.~Giannotti and E.~Mottola,
Phys. Rev. D \textbf{79} (2009), 045014
doi:10.1103/PhysRevD.79.045014
[arXiv:0812.0351 [hep-th]].
\bibitem{Armillis:2009pq}
R.~Armillis, C.~Coriano and L.~Delle Rose,
Phys. Rev. D \textbf{81} (2010), 085001
doi:10.1103/PhysRevD.81.085001
[arXiv:0910.3381 [hep-ph]].

\bibitem{Diaz:1989nx}
A.~Diaz, W.~Troost, P.~van Nieuwenhuizen and A.~Van Proeyen,
Int. J. Mod. Phys. A \textbf{4} (1989), 3959
doi:10.1142/S0217751X8900162X

\bibitem{Baikov:1991qz}
P.~A.~Baikov and V.~A.~Ilyin,
Theor. Math. Phys. \textbf{88} (1991), 789-809
doi:10.1007/BF01019107
\bibitem{Novotny:1994yx}
J.~Novotny,
Czech. J. Phys. \textbf{44} (1994), 633-661
doi:10.1007/BF01694837
\bibitem{Vassilevich:2003xt}
D.~V.~Vassilevich,
Phys. Rept. \textbf{388}, 279-360 (2003)
doi:10.1016/j.physrep.2003.09.002
[arXiv:hep-th/0306138 [hep-th]].
\bibitem{DeWitt:1964mxt}
B.~S.~DeWitt,
Conf. Proc. C \textbf{630701}, 585-820 (1964)
\bibitem{McKean:1967xf}
H.~P.~McKean and I.~M.~Singer,
J. Diff. Geom. \textbf{1}, 43-69 (1967)

\end{thebibliography}
\end{document}